# Metal Insulator Transition of Cr doped $V_2O_3$ calculated by hybrid density functional

Yuzheng Guo[1], Stewart J. Clark[2], and John Robertson[1*]

[1]Department of Engineering, University of Cambridge, Cambridge CB3 0FA, United Kingdom

[2]Department of Physics, Science Laboratories, University of Durham, South Road, Durham DH1 3LE, United Kingdom

Abstract: The electronic structure of vanadium sesquioxide in its different phases has been calculated using the screened exchange (sX) hybrid functional. The hybrid functional reproduces the electronic properties of all three phases, the paramagnetic metal (PM) phase, the anti-ferromagnetic insulating phase, and the Cr-doped paramagnetic insulating (PI) phase. A fully relaxed supercell model of Cr-doped $V_2O_3$ has a polaronic distortion around the substitutional Cr atoms and this local strain drives the PI-PM transition. The PI phase has a calculated band gap of 0.15eV in good agreement with experiment.

The metal insulator transition (MIT) is one of the most intriguing topics in condensed matter physics[1]. The electronic behavior of some materials can change greatly with a small change of external parameters. Vanadium sesquioxide $V_2O_3$ is a Mott-Hubbard insulator at low temperature, and it undergoes various MITs by varying the temperature, pressure or doping,[2-4] as in Fig 1 inset. However, we still lack a simple but accurate description of the precise nature of the electronic structure changes across these transitions. Above the Neel temperature of 170K, $V_2O_3$ has a paramagnetic metallic (PM) phase with the R3m corundum structure. Cr doping or applying pressure transforms PM $V_2O_3$ to a paramagnetic insulating (PI) phase with the same corundum structure. By lowering temperature, the PM and PI phases transform to a monoclinic antiferromagnetic insulating (AFI) phase[4]. This makes $V_2O_3$ a valuable test bed for studying the MIT.

The electronic structure of $V_2O_3$ has often been calculated by density functional theory (DFT). However, DFT does not give a band gap in either PI or AFI phases, because of its inability to correctly describe correlated d electrons[3, 5]. A band gap can be created for the AFI phase with the DFT+U method by introducing an on-site Coulomb interaction U for d electrons[3]. However, the PI phase is still beyond the scope of DFT+U as U is comparable in size to the relevant band width. Several more advanced methods have been applied to the DFT results. First, the GW approximation, gives a band gap for AFI[6]. The only method to correctly reproduce all three phases is dynamic mean field theory (DMFT)[7-9]. But DMFT cannot handle many atoms, so it cannot easily treat the geometry relaxation in large supercells as in a realistic model of Cr doped $V_2O_3$.

Recent data show that the MIT in $V_2O_3$ requires a better description of its electronic structure. It was previously believed that pressure and doping had the same effect on the MIT[1, 4]. Thus many DMFT studies used the experimental contraction of lattice parameter and the atomicl coordinates of the pure $V_2O_3$ primitive cell to simulate Cr doping[9, 10]. However, recent experiments show that the PI-PM transition due to pressure and Cr doping are not equivalent[11]. Thus, we must allow Cr doping to change internal atomic positions. Here, we present the first study of Cr-doped $V_2O_3$ with local atomic relaxations, by substituting Cr atoms into large $V_2O_3$ supercells and relaxing internal geometries variationally using a hybrid functional that also gives good band gaps.

Hybrid functionals are family of orbital-dependent exchange-correlation functionals where a portion of (non-local) Hartree-Fock exchange is mixed into the local exchange-correlation density functional. Hybrid functionals have been used to calculate single particle band structures of numerous transition metal oxides[12,13]. The HSE functional was used previously to study the MIT in pure $VO_2$ with changing the crystal structure[14], but based on experimental atomic coordinates. Here, we calculate all the three phases of $V_2O_3$ using the screened exchange (sX) hybrid functional[15,16]. The key point is that hybrid functionals are generalised Kohn-Sham functionals so we can use them variationally to relax local structures, and they are efficient enough to treat supercells of order 100 atoms. For $V_2O_3$ we find that the sX functional will reproduce the correct single-particle electronic structure in each phase without any fitting parameter.

The calculations are carried out with the CASTEP plane-wave pseudopotential code[17]. Norm-conserving pseudopotentials were generated by the OPIUM code[18]. The plane wave cut off energy is set to 780eV with a convergence of less than 0.01eV per atom. For primitive cells the Brillouin zone integrations use a 4×4×4 Monkhorst-Pack (MP) grid. An MP grid of 5×5×5 is used to calculate the density of states (DOS).

The largest supercell for Cr-doping has 120 atoms with 2 Cr atoms, equivalent to 4.2% Cr. Two Cr atoms in the supercell are necessary to model the MIT, as a single spin from one

Cr atom causes the whole system to transform to an AFI phase. In the PI phase, the Cr-doped supercell is completely relaxed with spin-degenerate sX functional. The residual force is less than 0.01 eV/A. Only Γ point is used for the 120-atom super cell geometry relaxation. A 3×3×1 MP grid is used for DOS calculation.

$V_2O_3$ shares the same corundum structure above 170K as the neighoring sesquioxides $Ti_2O_3$ and $Cr_2O_3$. The V atoms are surrounded by an octahedron of oxygen atoms with a trigonal distortion, leading to a splitting of 3d states into singly degenerate $a_{1g}$ and doubly degenerated $e^{\pi}_g$ bands. It was shown both by experiment and calculation that V has a mixed orbital occupation where both $a_{1g}$ and $e_g$ states are partially filled[3, 9, 10, 20].

Fig 1 shows the total DOS, partial DOS (PDOS) and band structure of the PM phase from a spin-unpolarised sX calculation using experimental lattice constants (a=4.951Å and c=14.004 Å). The Fermi level $E_F$ is set to 0 eV and lies within the V 3dπ states. The O 2p levels lie from -4eV to -8 eV below $E_F$ and these hybridize quite strongly with the V dσ states, consistent with photoemission[21] (Fig 2). The V 3dπ states at above −1 eV hybridize weakly with O p states. The total DOS is similar to that from LDA and DMFT[5, 9]. However, there is a weak peak at -0.8eV in the V 3d states, and the O 2p states are shifted down by 1eV compared to LDA, but consistent with photoemission[19,22] and DMFT[9]. We can compare the band structure with the crystal field theory of Goodenough[23]. There is no band gap between $a_{1g}$ and $e^{\pi}_g$ states and the bands are highly hybridized at L near $E_F$. The lowest band at -0.8 eV has a sizable dispersion along Γ-L-M representing the $a_{1g}$ bonding state of the V-V pair.

The corundum phase transforms to the AFI I2/a monoclinic phase below 150K. The phase transition is first-order with 8% volume decrease. The V atoms are ferromagnetically coupled in the (010) planes and anti-ferromagnetically coupled between the layers, as in experiment[26]. The internal geometry was relaxed by sX and the band structure calculated for this structure.

Fig 3 plots the total and PDOS for V and O in the AFI phase. The O 2p states are still separated from the V 3d states but the hybridization between O 2p and V 3d is stronger than in the PM phase. Both the valence band maximum (VBM) and the conduction band minimum (CBM) consists of V 3d states which confirms that AFI $V_2O_3$ is a Mott-Hubbard insulator. Compared to the PM phase, both the O 2p and V 3d states are broader (Fig 2). One sharp peak at -0.8eV in the PM phase disappears and several small peaks from 0 to -1.5eV appear, consistent with photoemission for the AFI phase[19]. A recent high resolution hard X-ray photoemission spectrum confirms that the quasi-particle spectrum below $E_F$ has many components[22] with two large peaks here lie at -0.3eV and -1.1eV, and our DOS has two large peaks at –0.1 eV and -1.2eV. LDA+DMFT cannot reproduce the first large peak near $E_F$[9]. The magnetic moment of the V atom is calculated to be 1.8 $\mu_B$, close to the LDA+U value[3] of 1.7 $\mu_B$. The spin is automatically relaxed in the spin-polarized calculation, thus this value identifies the S=1 local moment as ground state, consistent with experiment[26] and previous analysis[3] and contrary to the Castellini model.

Fig 3(b) shows the band structure of the AFI phase. The VBM is set to 0eV. The calculated band gap of 0.63eV agrees well with the experimnetal optical gap of 0.50-0.66eV [25, 27, 28]. $V_2O_3$ has an indirect gap with a VBM between R and T, and a CBM at Γ. The direct gap at Γ exceeds the indirect band gap by 0.1eV. The band gaps from different calculations and experiments are compared in Table 1. Our results are similar to the band structure from the U+GWA method but the band gap is more accurate and covers more of reciprocal space.

$V_2O_3$ transforms from PM to the PI phase with Cr doping. We use a corundum supercell of 120 atoms with 2 Cr atoms and a hexagonal unit cell. The internal coordinates are relaxed by sX. The Cr atom, surrounded by an oxygen octahedron, has the same local symmetry as V, but individual bond lengths now change. Due to the limit of supercell size we could not study the experimental MIT at a Cr content of ~1.5%. But we do find that *c* increases by 1.3% and *a* decreases by 0.5%, consistent with experimental lattice parameter data[20].

Fig 4 shows the total and PDOS for a PI $V_2O_3$ cell with 4.2% Cr doping. The V 3d PDOS resembles that of the AFI phase. The V 3d states range from 0eV to -2eV almost as twice wide as in the PM phase. The multi-peak feature remains, with two prominent peaks in the PDOS at a similar energy as in AFI phase. The band gap of 0.16eV is consistent with the photoemission value of 0.1eV[19, 22]. VBM is localized around Cr atoms while the CBM is consists of one flat band delocalized on V atoms. There is a similar band of 0.19eV separating the CBM from the bulk conduction bands. We also plot the partial DOS of Cr and V. The Cr PDOS is similar to that of V. The states from V and Cr are highly hybridized. The most significant peak from the Cr PDOS is about -1.8eV from the Fermi level, in good agreement with the small peak at -2eV seen by Mo et. al.[19]. The highly localized band edges from 3d states suggest the PI phase to be a typical Mott-Hubbard type.

Fig 2 compares the calculated PDOS of V 3d electrons from all the three phases with the latest photoemission data. The spectra are in excellent agreement with photoemission. The PI phase is more similar to the AFI phase rather than the PM phase. The 3d band widths of the PI and AFI phases below $E_F$ are clearly wider than PM.

The metal-oxygen bond lengths are summarized in Table 2. The V-O bond length decreases by ~2% in the AFI phase compared to PM due to volume shrinkage. The average bond length in the PI phase is the same as in PM. However the Cr-O bond length of 2.07 Å is 2% larger than the V-O bond length of 2.02 Å. Therefore Cr-doping introduces significant local strain[29]. We have stretched the primitive cell of corundum PM phase in different modes including uniaxial, biaxial, and uniform tension. None of these macroscopic strain modes can alone induce the MIT.

The key question is whether electron doping or structural distortion drives the PI-PM MIT. In order to separate these two effects, we calculated the band structure of unrelaxed Cr-doped $V_2O_3$ and the relaxed geometry without doping using the 120-atom supercells, as shown schematically in Fig. 5. Insulating cells are shown shaded. The relaxed 'undoped' cell is achieved by back-substituting V for Cr in the relaxed cell. The unrelaxed system remains metallic. The relaxed cell with no doping is insulating, with a similar band gap as the doped sample. Combined with the previous Cr PDOS analysis, we therefore argue that local strain not doping causes this MIT. If local strain is the driving force for the transition, this will depend on dopant positions. If the two Cr atoms are far away from each other as above, the supercell is found to be insulating. However if the two Cr atoms are together as a dimer, fig 5e, the effective strain is less and the cell remains metallic. The Cr-dimer doped $V_2O_3$ further confirms that electron doping is not the main driving force. Overall, the extra 2 electrons of two Cr atoms enter an extra valence state, and the MIT works by strain opening of the band gap, and not by Fermi level shifts in a fixed DOS.

We did not expect sX to match the photoemission spectrum because the dynamical effects included in DMFT are not present. But the band gap opening in different phases of $V_2O_3$ shows that the band structure of strong correlated electrons is greatly improved over LDA by introducing the non-local exchange. sX provides a good description of single particle states in strong-correlated systems at an affordable computational cost. Advanced methods such as DMFT and GW corrections can be applied on top of sX to include dynamic effects.

In summary, we have calculated the electronic structure for the PM, PI, and AFI phases of $V_2O_3$ with the sX hybrid functional. The PI phase is simulated by 4.2% Cr doping in a supercell, and the atomic geometry is relaxed with sX. The sX functional with local relaxations correctly gives the metallic and both insulating phases. A calculation of unrelaxed and undoped $V_2O_3$ supercells shows that local distortions around Cr sites rather than electron doping is the main driving force for dopant-induced MIT. The band gaps of the PI and AFI phases are 0.15eV and 0.63eV, in good agreement with experiment. Overall, this suggsts that hybrid functionals shouold provide a simple method to analyse complicated metal insulator transitions with largw polaronic coupling, such as manganites.

1   M. Imada, A. Fujimori, and Y. Tokura, Rev. Mod. Phys. **70**, 1039 (1998).

2   S. Lupi, et al, Nat. Commun. **1**, 105 (2010).

3   S. Y. Ezhov, V. I. Anisimov, D. I. Khomskii, and G. A. Sawatzky, Phys. Rev. Lett. **83**, 4136 (1999).

4   D. B. McWhan and J. P. Remeika, Phys. Rev. B **2**, 3734 (1970).

5   L. F. Mattheiss, J. Phys.: Condens. Matter **6**, 6477 (1994).

6   S. Kobayashi, Y. Nohara, S. Yamamoto, and T. Fujiwara, Phys. Rev. B **78**, 155112 (2008).

7   A. I. Poteryaev, J. M. Tomczak, S. Biermann, A. Georges, A. I. Lichtenstein, A. N. Rubtsov, T. Saha-Dasgupta, and O. K. Andersen, Phys. Rev. B **76**, 085127 (2007).

8   A. Georges, G. Kotliar, W. Krauth, and M. J. Rozenberg, Rev. Mod. Phys. **68**, 13 (1996).

9   K. Held, G. Keller, V. Eyert, D. Vollhardt, and V. I. Anisimov, Phys. Rev. Lett. **86**, 5345 (2001).

10  G. Keller, K. Held, V. Eyert, D. Vollhardt, and V. I. Anisimov, Phys. Rev. B **70**, 205116 (2004).

11  F. Rodolakis, et al, Phys. Rev. Lett. **104**, 047401 (2010).

12  F. Iori, M. Gatti, and A. Rubio, Physical Review B **85**, 115129.

13  Y. Guo, S. J. Clark, and J. Robertson, J. Phys.: Condens. Matter **24**, 325504.

14  V. Eyert, Phys. Rev. Lett. **107**, 016401 (2011).

15  D. M. Bylander and L. Kleinman, Phys. Rev. B **41**, 7868 (1990).

16  S. J. Clark and J. Robertson, Phys. Rev. B **82**, 085208 (2010).

17  M. D. Segall, P. J. D. Lindan, M. J. Probert, C. J. Pickard, P. J. Hasnip, S. J. Clark, and M. C. Payne, J. Phys. C **14**, 2717 (2002).

[18]  A. M. Rappe, K. M. Rabe, E. Kaxiras, and J. D. Joannopoulos, Phys. Rev. B **41**, 1227 (1990).

[19]  S. K. Mo, H. D. Kim, J. D. Denlinger, J. W. Allen, J. H. Park, A. Sekiyama, A. Yamasaki, S. Suga, Y. Saitoh, T. Muro, and P. Metcalf, Phys. Rev. B **74**, 165101 (2006).

[20]  F. Rodolakis, et al, Phys. Rev. B **84**, 245113 (2011).

[21]  E. Papalazarou, et al, Phys. Rev. B **80**, 155115 (2009).

[22]  H. Fujiwara, et al, Phys. Rev. B **84**, 075117 (2011).

[23]  J. B. Goodenough and F. A. Cotton eds., *Magnetism and the Chemical Bond* (Wiley, New York, 1963).

[24]  P. D. Dernier and M. Marezio, Phys. Rev. B **2**, 3771 (1970).

[25]  G. A. Thomas, D. H. Rapkine, S. A. Carter, A. J. Millis, T. F. Rosenbaum, P. Metcalf, and J. M. Honig, Phys. Rev. Lett. **73**, 1529 (1994).

[26]  R. M. Moon, Phys. Rev. Lett. **25**, 527 (1970).

[27]  M. M. Qazilbash, A. A. Schafgans, K. S. Burch, S. J. Yun, B. G. Chae, B. J. Kim, H. T. Kim, and D. N. Basov, Phys. Rev. B **77**, 115121 (2008).

[28]  G. A. Sawatzky and D. Post, Phys. Rev. B **20**, 1546 (1979).

[29]  A. I. Frenkel, D. M. Pease, J. I. Budnick, P. Metcalf, E. A. Stern, P. Shanthakumar, and T. Huang, Phys. Rev. Lett. **97**, 195502 (2006).

|  | Optical Gap[22, 25, 26] | LDA[5] | LDA+U[3] | GW[6] | GW@LDA+U[6] | sX |
|---|---|---|---|---|---|---|
| $E_g$(direct), eV | 0.5-0.66 | 0 | 0.60 | 0 | 0.943 | 0.76 |
| $E_g$(indirect), eV |  |  |  |  | 0.835 | 0.63 |

Table 1. Comparison of experimental bandgaps and those calculated by different functionals.

|  | PM corundum | AFI monoclinic | PI corundum |
|---|---|---|---|
| Metal-O bond length (Å) | 2.02 | 1.94 | 2.02(V-O) 2.07(Cr-O) |
| Metal-metal bond | 2.78 | 2.87 | 2.81(V-V) 2.80(Cr-V) |

Table 2. Comparison of metal-oxygen bond lengths in each phase.

Figure 1 (a) The total and partial DOS of PM phase. (b) The band structure of PM phase. Fermi level at 0eV. Inset - phase diagram of $V_2O_3$. Paramagnetic metal (PM) phase and paramagnetic insulating (PI) phase both have corundum structure while the anti-ferromagnetic insulating (AFI) phase is monoclinic.

Figure 2. Total DOS of each phase compared to photoemission spectra[19].

Figure 3 (a) The DOS and partial DOS of monoclinic AFI phase. (b) The band structure of monoclinic AFI phase. VBM at 0eV.

Figure 4 (a) The total and partial DOS of corundum PI phase.

Figure 5 (a) Perfect cell (b) Cr doped cell with no relaxation (c) Cr doped relaxed cell (d) Relaxed doped cell but without doping (e) Cr-dimer doped fully relaxed cell. Insulating phases shown shaded. *Note correlation of insulating to relaxed cell with separated dopants.*

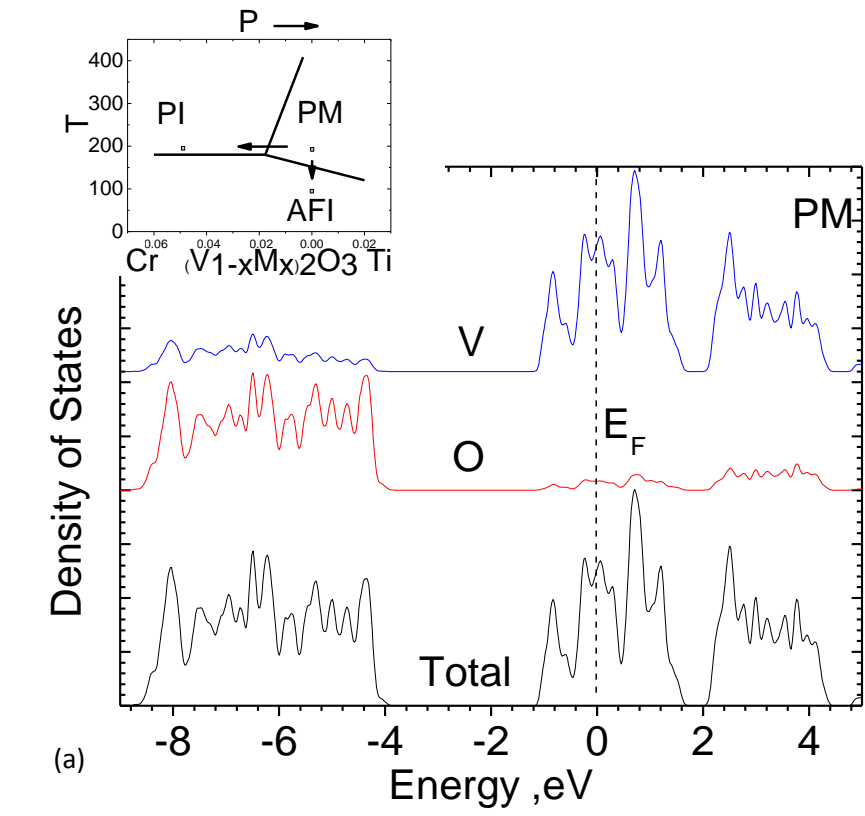

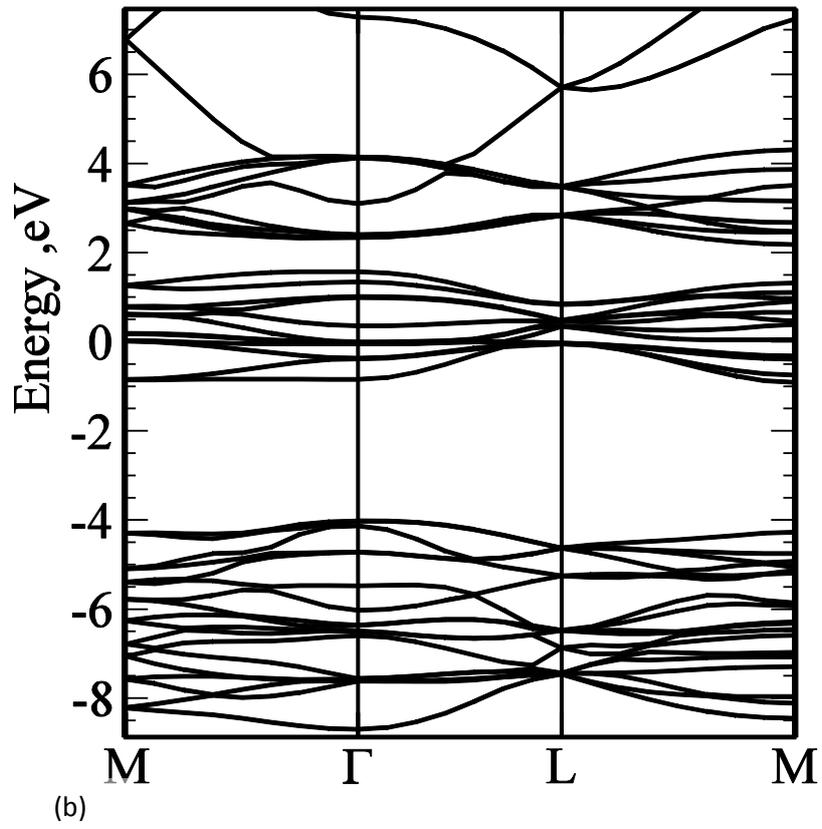

Fig .1

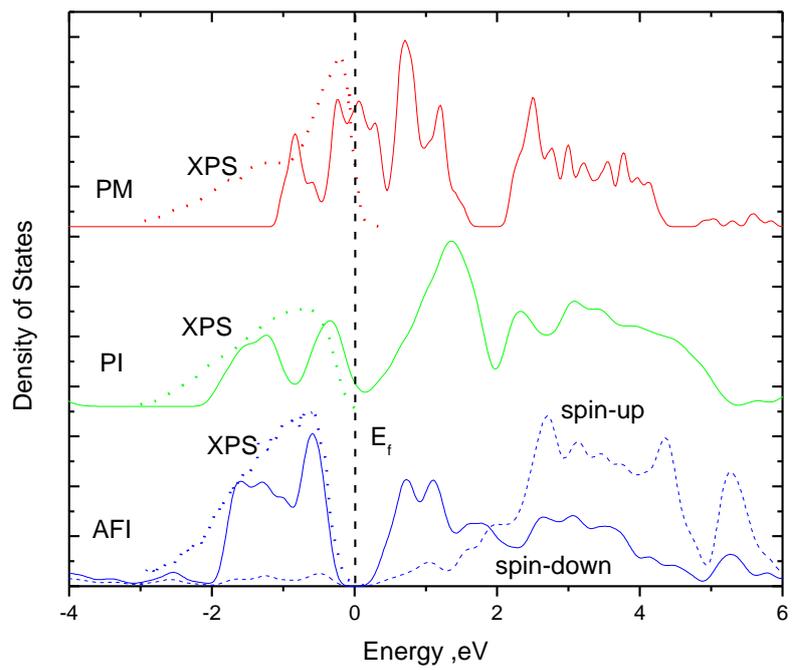

Fig 2

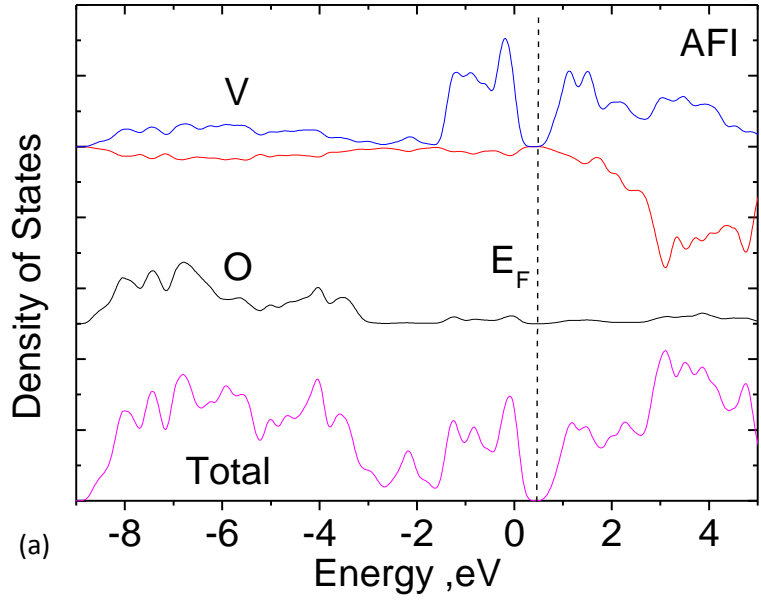

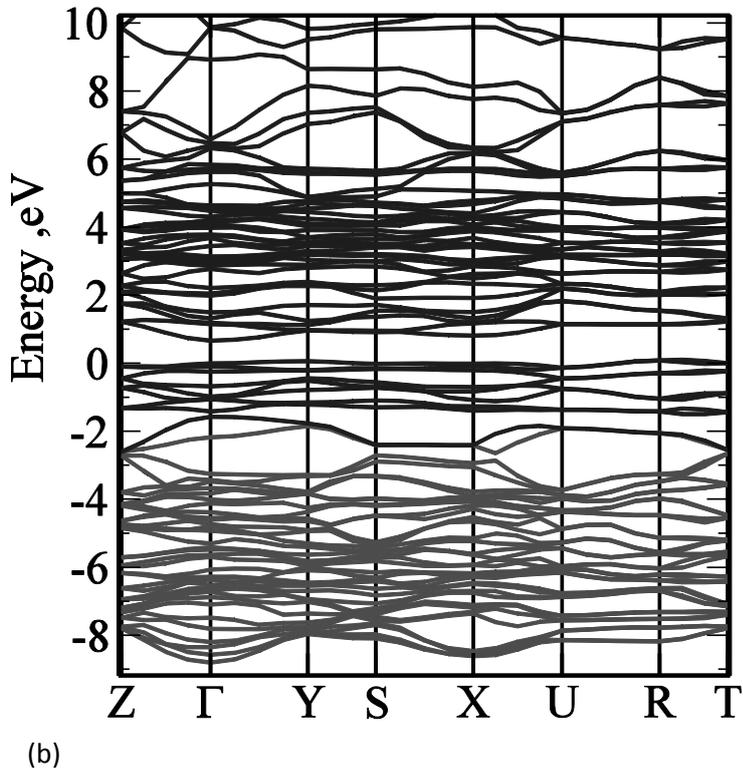

Fig. 3

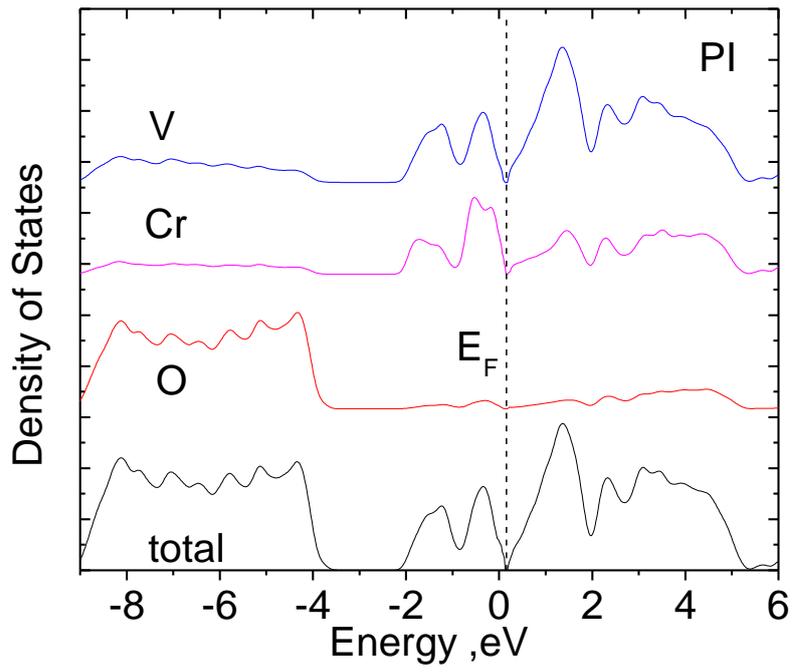

Fig. 4

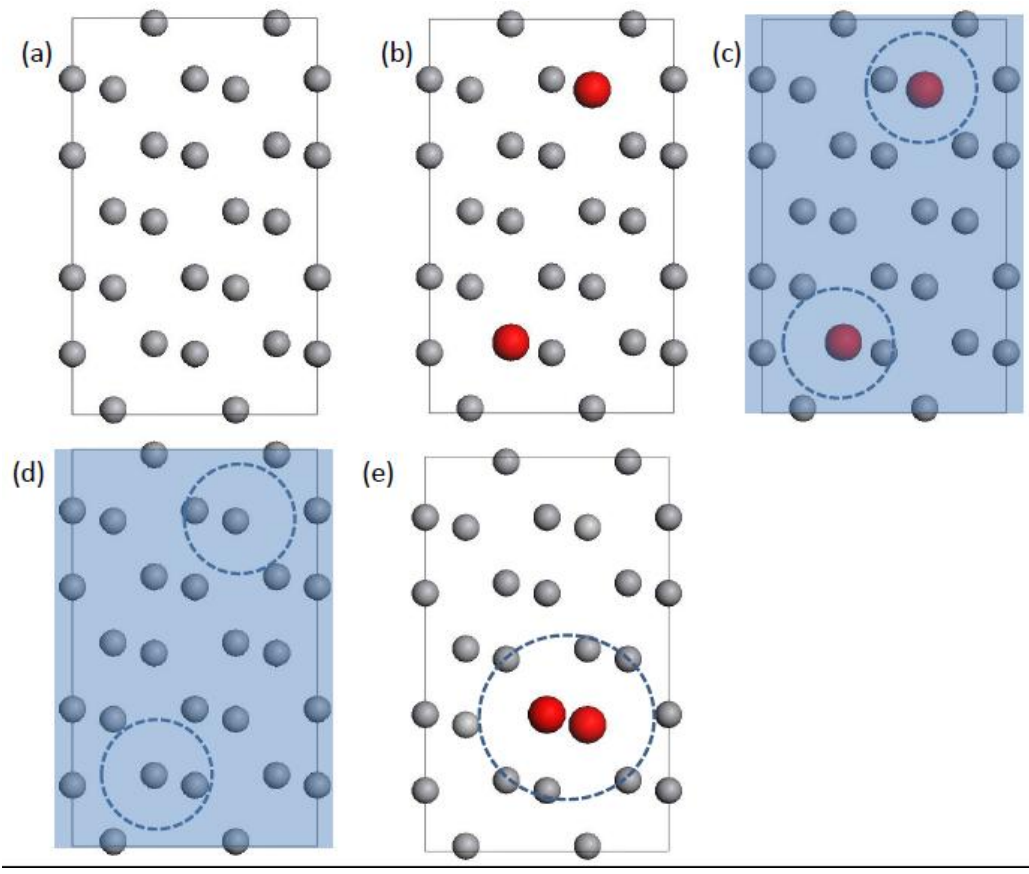

Fig 5